\documentclass[pre,amsmath,twocolumn,floatfix,showpacs]{revtex4}

\usepackage{graphicx}
\begin{document}

\title{A Novel Path Sampling Method for the Calculation of Rate
Constants}

\author{Titus S. van Erp, Daniele Moroni, and Peter G. Bolhuis}

\affiliation{Department of Chemical Engineering, Universiteit van
Amsterdam, Nieuwe Achtergracht 166, 1018 WV Amsterdam, The
Netherlands}

\date{\today}

\begin{abstract}
We derive a novel efficient scheme to measure the rate constant of
transitions between stable states separated by high free energy
barriers in a complex environment within the framework of transition
path sampling. The method is based on directly and simultaneously
measuring the fluxes through many phase space interfaces and increases
the efficiency with at least a factor of two with respect to existing
transition path sampling rate constant algorithms. The new algorithm
is illustrated on the isomerization of a diatomic molecule immersed in
a simple fluid.
\end{abstract}

\maketitle

\section{INTRODUCTION}
The calculation of rate constants of activated processes dominated by
 rare events, chemical reactions being a prime example, is still one
 of the major computational challenges.  As transition rates depend
 exponentially on the activation barrier height, the expectation time
 for an event can exceed current computer capabilities by many orders
 of magnitude.  As a result most chemical reactions can not be
 simulated by direct molecular dynamics (MD) methods, except those
 with very low activation energies.  The conventional way to tackle
 this time scale problem is based on transition state theory (TST) and
 separates the problem in two
 steps~\cite{Keck62,Anderson73,Bennet77,DC78}.  The first step is the
 calculation of the free energy barrier as function of a reaction
 coordinate, the second stage is the calculation of the transmission
 coefficient by sampling fleeting trajectories departing from the top
 of the barrier.  If the reaction coordinate is well chosen, the top
 of the free energy barrier corresponds to points in phase space close
 to the true transition state, and the transmission coefficient will
 have a reasonable value.  However, in high dimensional complex
 systems the choice of reaction coordinate can be extremely difficult
 and usually requires detailed {\em a priori} knowledge of the
 transition mechanism. Consequently, an intuitively chosen but wrong
 reaction coordinate can result in a very low transmission
 coefficient, and hence a statistically inaccurate or immeasurable
 rate constant.

Chandler and collaborators~\cite{TPS98,
TPS98a,TPS99,Bolhuis02,Dellago02} devised a method for which no prior
knowledge of the system is needed.  This method, called transition
path sampling (TPS), gathers a collection of trajectories connecting
the reactant to the product region by employing a Monte Carlo (MC)
algorithm. The resulting path ensemble can be used to elucidate
reaction mechanisms, transition states and reaction coordinates.  The
TPS method has been successfully used on such diverse systems as
cluster isomerization, auto-dissociation of water, ion pair
dissociation and on isomerization of a dipeptide, as well a reactions
in aqueous solution (see Ref.~\cite{Bolhuis02} for an overview).  Just
as in the conventional case mentioned above, an additional second
simulation is needed to determine the rate constant within TPS.  This
simulation combines the path sampling method with the umbrella
sampling technique to estimate the probability to reach the product
state from the initial reactant state.  The final macroscopic rate
constant is given by a plateau in the time derivative of a correlation
function~\cite{TPS99}.  In case of two distinct stable states 
this plateau region should always exist at times longer
than the typical molecular relaxation time.  However, when reaction
pathways are complex and exhibit multiple recrossings, these typical
molecular relaxation timescales can be relatively long.  In that case
the TPS rate constant calculation is computationally expensive, as the
path length must exceed these timescales.

In this paper we improve the efficiency of the TPS rate constant
calculation on several points by introducing an alternative scheme for
calculating reaction rates, named transition interface sampling (TIS).
The first of these improvements is allowing the path length to vary,
so that by a well chosen definition of the stable states we can limit
the length of each path to the strict minimum.  Secondly, the new
method is based on the effective positive flux through dividing
surfaces or interfaces and is consequently much less sensitive to
multiple recrossings or diffusive barrier crossings.  Thirdly, the
number of different types of Monte Carlo moves is reduced, making the
implementation of the algorithm conceptually simpler.

This paper is organized as follows: In Sec.~\ref{theory} we briefly 
describe the existing algorithms and present
the theoretical derivation for the TIS rate constant expression.  The
implementation of the algorithm is discussed in Sec.~\ref{algorithm}.
We illustrate the algorithm on a diatomic molecule in a fluid of
repulsive particles and make a quantitative comparison to the original
TPS calculation in Sec.~\ref{application}. We end with concluding
remarks in Sec.~\ref{conclusion}.

\section{THEORY}
\label{theory}

\subsection{Transition state theory and the calculation of rate
constants}

Consider a dynamical system in which transitions can take place
between two stable states $A$ and $B$.  If the barrier between $A$ and
$B$ is sufficiently high, the system will show exponential relaxation
for which the forward and backward rate constants $k_{AB}$ and
$k_{BA}$ are well defined and can be expressed in terms of microscopic
properties.  Measuring these rate constants by computer simulation is
traditionally done by the two stage Bennett-Chandler (BC) procedure
based on the principles of 
TST~\cite{Bennet77,DC78}.  The first step is the calculation of the
reversible work or free energy to bring the system from stable state
$A$ to the transition state. This free energy $F(\lambda)$ has to be
calculated as a function of a suitably chosen reaction coordinate
$\lambda$. This $\lambda$ can be a complex function of all particle
coordinates $r$ and momenta $p$: $\lambda=\lambda(x)$, with $x\equiv
\{r,p\}$.  The maximum in $F(\lambda)$ defines the transition state
dividing surface $\lambda^*$~\cite{FrenkelSmit,Chandlerbook}.  By
convention, the system is in $A$ if $\lambda(x) < \lambda^*$ and in
$B$ if $\lambda(x) > \lambda^*$.

The main assumption in TST is that any trajectory coming from $A$ and
crossing the transition state dividing surface $\lambda(x)=\lambda^*$
will remain at the $B$ side of the dividing surface for a long time.
The reaction rate can therefore be expressed as the positive flux
through the multidimensional dividing surface $\lambda^*$
\begin{eqnarray}
\label{rateflux}
k_{AB}^{TST} & = & \lim_{\Delta t \rightarrow 0} \frac{1}{\Delta t}
\frac{ \left \langle \theta\big(\lambda^*-\lambda(x_0) \big)
\theta\big( \lambda(x_{\Delta t}) -\lambda^* \big) \right \rangle }{
\left \langle \theta\big( \lambda^*-\lambda \big) \right \rangle }
\nonumber\\ & = & \frac{ \left \langle \dot{\lambda}(x_0) \delta\big(
\lambda(x_0) -\lambda^* \big) \theta\big( \dot{\lambda}(x_0) \big)
\right \rangle }{ \left \langle \theta\big( \lambda^*-\lambda \big)
\right \rangle }\nonumber\\ & = & \left \langle \dot{\lambda}(x_0)
\theta\big( \dot{\lambda}(x_0) \big) \right \rangle_{\lambda^*}
\frac{e^{-\beta F(\lambda^*)}}{\int_{-\infty}^{\lambda^*} e^{-\beta
F(\lambda)} d\lambda},
\end{eqnarray}
where $x_t$ specifies the set of coordinates and momenta of the system
at time $t$, the dots denote derivatives with respect to time $t$, the
brackets $\left \langle \ldots \right \rangle$ denote equilibrium
ensemble averages and $\theta(x)$ and $\delta(x)$ are the Heaviside
step-function and the Dirac delta function respectively.  In the last
equality of Eq.~(\ref{rateflux}) the connection to the reversible work
$F(\lambda)$ is made, and $\beta=1/k_{\rm B} T$, where $k_{\rm
B}$ is Boltzmann's constant and $T$ is the temperature.  The subscript
$\lambda^*$ to the ensemble brackets, indicates that the ensemble is
constrained to the top of the barrier $\lambda^*$.

We consider the system to be completely deterministic and thus we can
write $x_t=f(x_{t'},t-t')=f(x_0,t)$, in which $f$ is the
time-propagator function.  Evaluation of the function $f(x,t)$
requires integrating the equations of motion over the time interval
$t$ starting with configuration $x$.  Nevertheless, the equations
derived in this paper are still valid when applied to stochastic
dynamics.

\begin{figure}
\includegraphics[width=8cm]{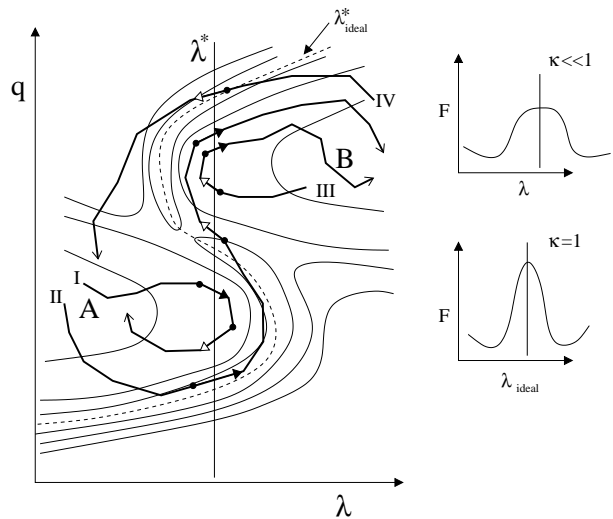}
\caption{The thin solid curve show the two dimensional free energy
landscape in contour plot. $\lambda$ is the chosen reaction
coordinate, $q$ represents all other degrees of freedom.  $A$ and $B$
denote the state regions.  The vertical line at $\lambda^*$
corresponds to the maximum in the free energy function $F(\lambda)$ as
is shown at the right upper side.  The free energy as function of the
ideal reaction coordinate is also shown at the right lower side.  This
reaction coordinate is a complex function of all degrees of freedom
$\lambda_{\textrm{ideal}}=\lambda_{\textrm{ideal}}(\lambda,q)$ and the
corresponding free energy function has its maximum at the true
transition state dividing surface
$\lambda_{\textrm{ideal}}=\lambda_{\textrm{ideal}}^*$.  This true
dividing surface is the dashed curved line.  The corresponding free
energy barrier is much more narrow and higher than the artificial
barrier due to the incorrect reaction coordinate. Four possible
trajectories are shown. The black solid arrows indicate a positive
flux through the surface $\lambda^*$ and the white solid arrows
indicate the negative fluxes. TST rate expression~(\ref{rateflux})
counts all positive fluxes of trajectories I, II and
III. Consequently, non-true reactive events like I and III have a
artificial contribution to the rate constant and also trajectory II is
overcounted one time.  To correct for this, one can calculate the
transmission coefficient $\kappa$.  In the TPS
equation~(\ref{rateTPS}), if $\lambda_A=\lambda_B=\lambda^*$,
trajectories III and IV are not counted because of the $h_A(x_0)$
term.  Trajectories I and II are correctly counted in the final
summation due to the cancellation of positive and negative flux
terms.\label{trajcount}}
\end{figure}

Even when the TST assumption is accurate, it can be extremely
difficult to find a proper reaction coordinate for which recrossings
do not occur. As a result a wrong choice for the reaction coordinate
will give a much lower free energy barrier than the real activation
free energy and will correspondingly overestimate the rate constant.
Figure~\ref{trajcount} illustrates that Eq.~(\ref{rateflux}) overcounts
trajectories. One can correct for this overcounting by multiplying the
TST rate constant with the transmission coefficient $\kappa(t)$ to
obtain the true rate constant
\begin{equation}
k_{AB}(t) = k^{TST}_{AB} \kappa (t).
\end{equation}

The calculation of the time dependent transmission coefficient
$\kappa(t)$ constitutes the second part of the two stage BC
procedure~\cite{Bennet77,DC78}.  $\kappa(t)$ belongs to the
approximate dividing surface
$\lambda^*$~\cite{Keck62,Anderson73,Bennet77,DC78} and can be
determined by taking an ensemble average of many short trajectories
starting from the dividing surface:
\begin{eqnarray}
\kappa(t) & = & \frac{1}{ \left \langle \dot{\lambda}(x_0) \theta\big(
\dot{\lambda}(x_0) \big) \right \rangle_{\lambda^*}} 
\left \langle \dot{\lambda}(x_0)\theta\big( \lambda (x_t) -\lambda^*
\big) \right \rangle_{\lambda^*}.
\end{eqnarray}
After a short molecular time $t_{\rm mol}$ the trajectories are
committed to a stable state and $\kappa(t)$, and hence $k_{AB}(t)$,
become constant: the transmission coefficient $\kappa$, and the rate
constant $k_{AB}$, respectively.  It is however important to start
sufficiently close to the true transition state dividing surface.
Otherwise the transmission coefficient will be extremely low, making
an accurate estimate of the rate constant problematic or even
impossible.  In many cases, in particular for complex condensed matter
systems, a sufficiently close reaction coordinate is difficult to find
and requires considerable a-priori knowledge about the system.

\subsection{Transition path sampling}
\label{sec.TPS}

Transition path sampling (TPS) is developed to overcome the
difficulties mentioned above
\cite{TPS98,TPS98a,TPS99,Bolhuis02,Dellago02}.  Its main advantage is
that no prior knowledge of the transition state is needed.  The rate
constant in TPS is expressed as the time derivative of a general time
correlation function.

\begin{equation} k_{AB}^{TPS}(t)=
\frac{\mathrm{d}}{\mathrm{d}t} C(t), \quad C(t)= \frac{ \left \langle
h_A(x_0) h_B(x_t) \right \rangle }{\left \langle h_A(x_0) \right
\rangle },
\label{rateTPS}
\end{equation}
in which $h_A(x)$ and $h_B(x)$ are the characteristic functions
defined by:
\begin{eqnarray}
h_A(x) &=& 1, \quad \textrm{ if } x \in A, \quad \textrm{ else } \quad
h_A(x)=0 \nonumber \\ h_B(x) &=&1, \quad \textrm{ if } x \in B, \quad
\textrm{ else } \quad h_B(x)=0.
\label{charAB}
\end{eqnarray}

In case of a single order parameter $\lambda(x_t)$ describing the
transition, the phase space regions $A$ and $B$ are defined by
$\lambda_A$ and $\lambda_B$: $x_t \in A \textrm{ if } \lambda ( x_t )
< \lambda_A$ and $x_t \in B \textrm{ if } \lambda(x_t) > \lambda_B$.
Knowledge of the precise location of the dividing surface $\lambda^*$,
$\lambda_A < \lambda^* < \lambda_B$, is not required in TPS.
Therefore, the order parameter $\lambda$ does usually not correspond
to the reaction coordinate.

The microscopic expression for the rate constant in
Eq.~(\ref{rateTPS}) is time dependent, while the phenomenological rate
constant is not. However, just as the transmission coefficient
$\kappa(t)$ becomes a constant, the time dependent function $
k_{AB}^{TPS}(t)$ reaches a plateau after a molecular timescale $t_{\rm
mol}$. The phenomenological rate constant is equal to the plateau
value: $k_{AB}=k_{AB}^{TPS}(T)$.  This plateau region should always
exist for times $T$ between the molecular timescale and the
characteristic reaction time: $t_{\rm{mol}} < T \ll t_{\rm{rxn}}$. In
other words, $T$ is larger than the timescale to commit to one of the
stable states, but much shorter than the expectation time
$t_{\rm{rxn}}$ of a completely new reactive event.  If we take
$\lambda_A=\lambda_B=\lambda^*$ and the limit $t \rightarrow 0+$,
Eq.~(\ref{rateTPS}) transforms into the expression for the positive
reactive flux or, equivalently, the TST equation (\ref{rateflux}).
For $t>0$, however, the reactive flux measured by Eq.~(\ref{rateTPS})
no longer consists of purely positive contributions. The final rate
constant is a sum of positive and negative fluxes, and thus the
overcounting of trajectories in Eq.~(\ref{rateflux}) is
circumvented. (See fig.(\ref{trajcount})).

We can rewrite the time dependent rate constant of Eq.~(\ref{rateTPS})
into~\cite{TPS99}:
\begin{equation}\label{kTPS}
k_{AB}^{TPS}(t)=\frac{\langle\dot{h}_B(t)\rangle_{A,H_B(T)}}{\langle
h_B(t')\rangle_{A,H_B(T)}} \cdot C(t'),
\end{equation} 
where $H_B(T) =\max_{0<t<T} h_B(x_t)$ and $\langle \ldots
\rangle_{A,H_B(T)}$ denotes an average on the ensemble of paths of
fixed length $T$ starting in $A$ and entering $B$ at least
once~\cite{TPS99}. These ensemble averages are evaluated using a Monte
Carlo procedure employing the \emph{shooting} and \emph{shifting}
moves~\cite{TPS98a}.  The two factors in Eq.~(\ref{kTPS}) have to be
evaluated separately.  First, a path sampling simulation is performed
to compute $\langle h_B(t)\rangle_{A,H_B(T)}$ in the interval $[0,T]$.
The path length $T$ must be long enough for the time derivative to
display a plateau.  Subsequently, one chooses a $t'$ in interval
$[0,T]$ and computes $C(t')$ using the path sampling in combination
with an umbrella sampling technique~\cite{TPS99}.  A drawback of the
TPS rate constant calculation is that the function $k_{AB}^{TPS}(t)$
can be strongly oscillatory because of recrossings and will reach a
plateau only after a relatively long time.  The path length in TPS
must exceed the typical timescale of these oscillations, and
consequently, in that case TPS is computationally costly.

\subsection{Transition Interface Sampling}
\label{sec.TIS}

Just as the BC and the TPS rate constant algorithms, the TIS method is
based on a flux calculation. In contrast to these schemes, however, TIS
measures the effective positive flux~\footnote{Here, effective means
that the recrossings through the interfaces are not being counted},
instead of a conditional general flux as in Eq.~(\ref{rateTPS}) or
Eq.~(\ref{rateflux}).  This implies that only positive terms
contribute to the rate, allowing for faster numerical convergence.  A
flux is normally defined through a hypersurface in phase space defined
by an order parameter, the reaction coordinate.  But, similar to the
TPS case, we do not want to suffer from a bad choice of reaction
coordinate. Therefore, instead of using a single dividing surface, we
introduce a series of interfaces through which we measure this flux.
We then derive an expression that relates the flux through a certain
interface to the flux through an interface which is closer to $A$ to
replace the expensive TPS umbrella sampling procedure.

In order to formulate a proper flux, we have to divide the entire
phase space into two complementary regions called  {\em overall}
states $\mathcal{A}$ and $\mathcal{B}$. These states do not only
depend on the position at the time of consideration but also on its
past behavior.  Overall state $\mathcal{A}$ covers all phase space
points lying inside stable region $A$, which constitutes the largest
part, but also all phase space points that visit $A$, before reaching
$B$ when the equations of motion are integrated backward in time.
Similarly, state $\mathcal{B}$ comprises stable state $B$ and all
phase points, coming directly from this state in the past,
i.e. without having been in $A$.  It is useful to generalize the
characteristic functions in Eq.(\ref{charAB}) for an arbitrary phase
space region $\Omega$
\begin{eqnarray}
h_\Omega(x) &=& 1, \quad \textrm{ if } x \in \Omega, \quad \textrm{
else } \quad h_\Omega(x)=0.
\end{eqnarray}
For each phase point $x$ and each phase space region $\Omega$ we can
determine the minimum (first entrance) times $t_\Omega^b(x)$ and
$t_\Omega^f(x)$ needed to reach $\Omega$ starting from configuration
$x$ by integrating the equations of motion backward and forward in
time, respectively:
\begin{eqnarray}
t_\Omega^b(x) &\equiv& -\max \left[ \{ t | h_\Omega(f(x,t))=1 \wedge t
\leq 0 \} \right] \nonumber \\ t_\Omega^f(x) &\equiv& +\min \left[ \{
t | h_\Omega(f(x,t))=1 \wedge t \ge 0 \} \right],
\end{eqnarray}
where the $\min$ and $\max$ function return respectively the lowest
and highest value of their arguments.  In addition, it is useful to
define for each phase point $x$ and each set of two non-overlapping
phase space regions $\{\Omega_1,\Omega_2 \}$ the following
characteristic functions:
\begin{eqnarray}
\bar{h}_{\Omega_1,\Omega_2}^b(x) & = &
\begin{cases}
1 \quad \textrm{ if } h_{\Omega_1} \Big(f(x,-t_{\Omega_1 \cup
\Omega_2}^b(x) ) \Big)=1,\\ 0 \quad \textrm{ otherwise}
\end{cases} \nonumber \\
\bar{h}_{\Omega_1,\Omega_2}^f(x) & = &
\begin{cases}
1 \quad \textrm{ if } h_{\Omega_1} \Big(f(x,+t_{\Omega_1 \cup
\Omega_2}^f(x) ) \Big)=1,\\ 0 \quad \textrm{ otherwise}
\end{cases}
\end{eqnarray}
In words, these functions measure whether a trajectory reaches
$\Omega_1$ before $\Omega_2$ or not.  As the system is ergodic, each
phase space region will be visited in finite time and thus
$\bar{h}_{\Omega_1,\Omega_2}^b(x)+\bar{h}_{\Omega_2,\Omega_1}^b(x)=
\bar{h}_{\Omega_1,\Omega_2}^f(x)+\bar{h}_{\Omega_2,\Omega_1}^f(x)=1$
for any $x$.  Using these definitions the characteristic functions for
the overall states ${\mathcal A}$ and ${\mathcal B}$ are given by
\begin{eqnarray}\label{TISstates}
h_{\mathcal A}(x)=\bar{h}_{A,B}^b(x), \quad h_{\mathcal
B}(x)=\bar{h}_{B,A}^b(x).
\end{eqnarray}
These states together span the complete phase space, as the system can
never stay in the intermediate region between $A$ and $B$ forever.
The overall states ${\mathcal A}$ and ${\mathcal B}$ do not
sensitively depend on the definition of stable state $A$ and $B$ as
long as it is reasonably.  Of course, the stable regions should not
overlap, each trajectory between the stable states must be a true rare
event for the reaction we are interested in. In addition, the
probability that after this event the reverse reaction occurs shortly
thereafter must be as unlikely as an entirely new event. In other
words, the system must be committed to the stable states. Therefore, a
reasonable definition of $A$ and $B$ requires that they should lie
completely inside the basin of attraction of the respective two 
states~\footnote{In general, $A$ and $B$
are defined in phase space, but for most practical cases configuration
space might be enough.} (see also Ref.~\cite{Dellago02}). 
Special care has to be taken with this
condition for processes which show many recrossings between state $A$
and $B$ before settling down.  Such processes can occur in solution or
in dilute gasses.  For instance, for organic reactions in aqueous
solution, a rare specific hydrogen
bonded network can lower the bond-breaking barrier and initiate the
reaction.  If the lifetime of those rare solvation structures is high,
a sudden reverse reaction can occur as the barrier for the backward
reaction is also lowered by the same amount~\cite{Strnad,Okuno,Erp}.
A similar phenomenon can happen in dilute gasses for which rare
spontaneous fluctuations in the kinetic energy are the main driving
force.  A particle moving from one state to another due to a very high
kinetic energy as result of sequence of collisions can cross the
potential energy barrier several times before it will dissipate its
energy by a new collision and relax into one of the stable states (see
e.g. Refs.~\cite{CHO,Sun}).  These problems can in principle be solved
by an adequate choice of the stable state definitions.  For instance,
the definition can depend explicitly on the presence of certain
hydrogen bonds or on kinetic energy terms.

\begin{figure}
\includegraphics[width=8cm]{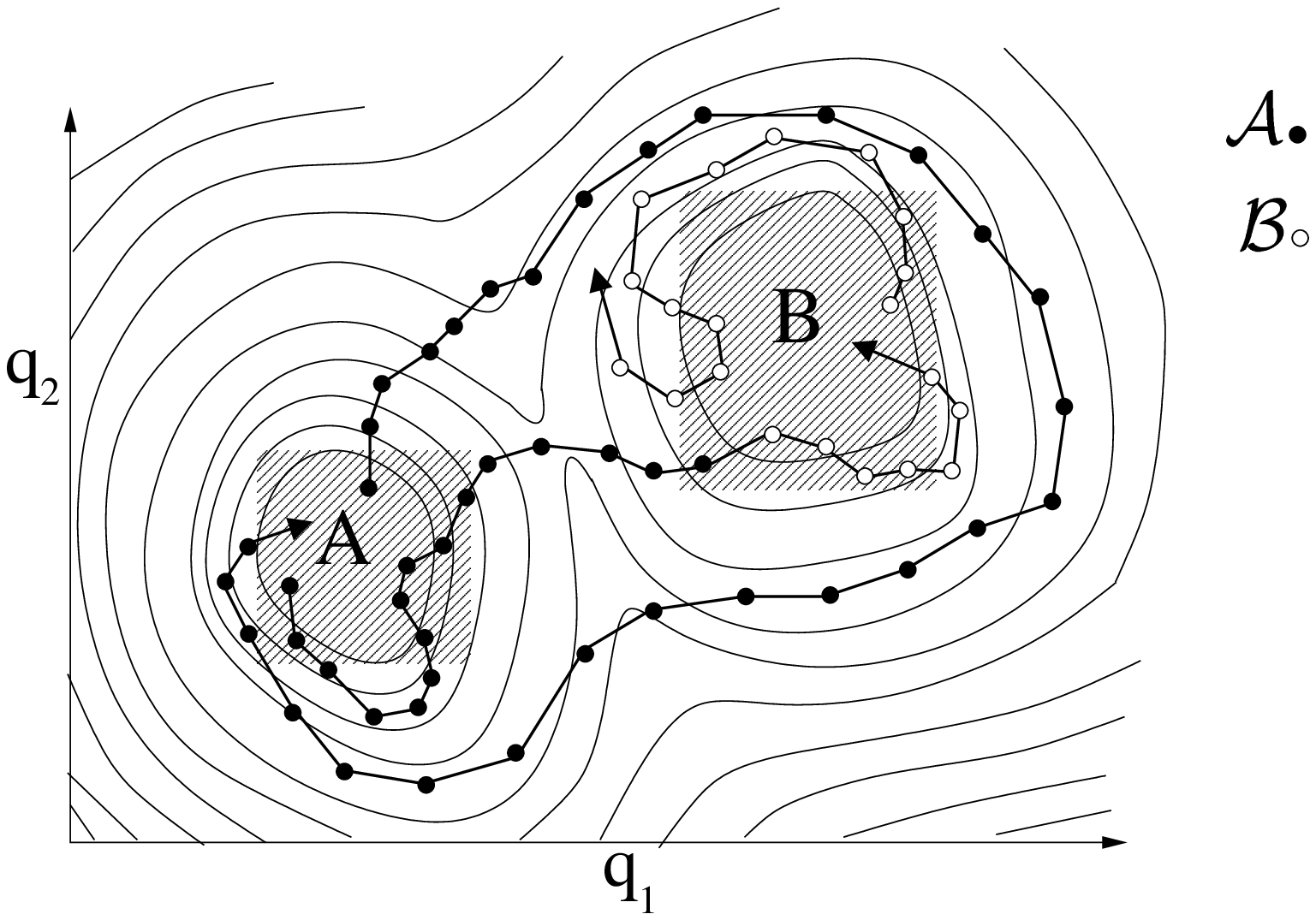}
\caption{Example of phase space regions in TIS. Thin solid curves
denote the free energy contour lines. $q_1$ and $q_2$ are two
arbitrary projections of the degrees of freedom. $A$ and $B$ are the
two stable states.  The dots on the three shown trajectories indicate
the positions of the system at successive time steps.  The overall
state $\mathcal{A}$ and $\mathcal{B}$ are indicated by black and white
dots respectively. Only one trajectory starts in $A$ and ends in $B$
and is therefore a true reactive event. The system changes from state
$\mathcal{A}$ into $\mathcal{B}$ when it enters region $B$ for the
first time. It can leave stable region $B$ shortly thereafter, but
never go back to $A$ in a short time. The stable regions have to be
chosen to fulfill that condition.} \label{stateAB}
\end{figure}

\begin{figure}
\includegraphics[width=8cm]{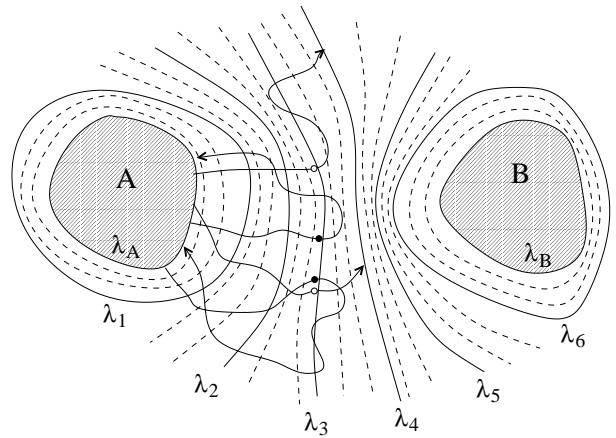}
\caption{Example of the division of the phase space by interfaces. $A$
and $B$ are the stable state regions with interfaces $\lambda_A$ and
$\lambda_B$. The interfaces $\lambda_1 \ldots \lambda_6$ correspond to
a calculation of Eq.~(\ref{rate4}) with $n=6$. The dashed lines are
the sub interfaces in between. Four trajectories are shown
corresponding to a ${\mathcal P}(\lambda_4|\lambda_3)$ ensemble
calculation. On each trajectory the $x_0$ time slice is indicated with
a circle. Black circles correspond to $
\bar{h}_{A,\Omega_{\lambda_4}}^f(x_0)=0$ and white circles correspond
to $ \bar{h}_{A,\Omega_{\lambda_4}}^f(x_0)=1$.} \label{interfaces}
\end{figure}

With our definition of overall states ${\mathcal A}$ and ${\mathcal
B}$ we can write down our rate equation in the spirit of
Eq.~(\ref{rateTPS}):
\begin{equation}
k_{AB}= \frac{\left \langle h_{\mathcal A}( x_0 ) \dot{h}_{\mathcal
B}( x_0 ) \right \rangle }{ \left \langle h_{\mathcal A}(x_0) \right
\rangle},
\label{rateTIS}
\end{equation}
where the dot denotes the time derivative taken at $t=0$.  This rate
expression does not depend on time although the evaluation of the
characteristic functions still requires integration of the equations
of motion.  The transition from $\mathcal{A}$ into $\mathcal{B}$ takes
place when the system coming from $A$ will cross the interface
$\lambda_B$ for the first time (see Fig.~\ref{stateAB}). After this
event the system will stay in $\mathcal{B}$. Eq.~(\ref{rateTIS})
counts therefore only the first crossing through interface $\lambda_B$
and is hence equivalent to the effective positive flux expression
\begin{eqnarray}
\label{rate_posflux}
k_{AB} & = & \frac{\left \langle h_{\mathcal A}(x_0)
\dot{\lambda}(x_0) \theta(\dot{\lambda}(x_0))
\delta(\lambda(x_0)-\lambda_B) \right \rangle}{ \left \langle
h_{\mathcal A}(x_0) \right \rangle} \\ & = & \lim_{\Delta t
\rightarrow 0} \frac{1}{\Delta t} \frac{\left \langle h_{\mathcal
A}(x_0) \theta(\lambda_B-\lambda(x_0) ) \theta(\lambda(x_{\Delta
t})-\lambda_B) \right \rangle}{ \left \langle h_{\mathcal A}(x_0)
\right \rangle}.\nonumber
\end{eqnarray}
Note the similarity with Eq.~(\ref{rateflux}).  Strictly speaking
$\theta(\lambda_B-\lambda(x_0))$ is redundant in
Eq.~(\ref{rate_posflux}) as $h_{\mathcal A}(x_0) =0$ if
$\theta(\lambda_B-\lambda(x_0)) \neq 1$.  The last expression in
Eq.~(\ref{rate_posflux}) is most suitable for a numerical approach
with $\Delta t$ as the time step in a molecular dynamics simulation.
Evaluation of Eq.~(\ref{rate_posflux}) requires counting all phase space
points which at $t=0$ are just about to cross interface $\lambda_B$ 
in one time step {\em and} will
enter region $A$ before $B$ when integrating backward in time starting
from $x_0$.  Unfortunately, Eq.~(\ref{rate_posflux}) is not very
efficient from a computational point of view because only a very small
fraction of phase points close to interface $\lambda_B$ actually
belong to $\mathcal{A}$, leading to poor statistics.  We can enhance
the statistical accuracy by relating the flux through $\lambda_B$ to
the flux through an interface closer to $A$. We therefore introduce a
set of $n$ non-intersecting interfaces $\lambda_1, \lambda_2,
\lambda_3, \ldots \lambda_n$, each interface $\lambda_i$ closer to $A$
than the next interface $\lambda_{i+1}$ (see
Fig.~\ref{interfaces}). We define the corresponding phase space
regions $\Omega_{\lambda_i} \equiv \{x| \lambda(x) > \lambda_i\}$.  In
this way $\Omega_{\lambda_B}$ is equivalent to our stable state $B$,
while $\Omega_{\lambda_A}$ is the phase space outside stable state
$A$.  By introducing the following definition
\begin{eqnarray}
\Phi_{A,\lambda_i}(x_0) \equiv \quad \quad \quad \quad \quad \quad
\quad \quad \quad \quad \quad \quad \quad \quad \quad \quad \nonumber
\\ \lim_{\Delta t \rightarrow 0} \frac{1}{\Delta t}
\bar{h}_{A,\Omega_{\lambda_i}}^b(x_0) \,
\theta(\lambda_i-\lambda(x_0)) \theta(\lambda(x_{\Delta
t})-\lambda_i)),
\end{eqnarray}
Eq.~(\ref{rate_posflux}) reduces to
\begin{equation}
\label{eq:kab_phi}
k_{AB}= \left \langle \Phi_{A,\lambda_B} \right \rangle / \left
\langle h_{\mathcal A} \right \rangle .
\end{equation}
where $\langle \Phi_{A,\lambda_i} \rangle$ denotes the effective
positive flux through interface $\lambda_i$.  The rate constant is
thus equal to the effective positive flux through interface
$\lambda_B$ with the condition the trajectories came directly from
$A$.  Note again that $\left \langle \Phi_{A,\lambda_A} \right \rangle
/ \left \langle h_{\mathcal A} \right \rangle $ is equal to the TST
rate expression in Eq.~(\ref{rateflux}) in case
$\lambda^*=\lambda_A=\lambda_B$.  The effective flux $ \langle
\Phi_{A,\lambda_i} \rangle $ can now be related to the effective flux
$ \langle \Phi_{A,\lambda_{i-1}} \rangle $ through an interface
$\lambda_{i-1}$ closer to $A$ by (see Appendix A)
\begin{equation} 
\langle \Phi_{A,\lambda_i}(x_0) \rangle = \left \langle
\bar{h}_{\Omega_{\lambda_i},A}^f(x_0) \right
\rangle_{\Phi_{A,\lambda_{i-1}}} \times \left \langle
\Phi_{A,\lambda_{i-1}}(x_0) \right \rangle,
\label{eq:recursive}
\end{equation}
where $\left \langle \ldots \right \rangle_{\Phi_{A,\lambda_{i-1}}}$
denotes the ensemble average over all phase space points $x_0$ for
which $\Phi_{A,\lambda_{i-1}}(x_0) \neq 0$.  The factor $ \left<
\right.  \bar{h}_{\Omega_{\lambda_{i}},A(x_0)}^f \left. \right
\rangle_{\Phi_{A,\lambda_{i-1}}} \equiv {\mathcal P}(\lambda_{i} |
\lambda_{i-1}) $ is the conditional probability that a trajectory,
coming from $A$, passes $\lambda_{i}$, given the fact that it has
passed the interface $\lambda_{i-1}$ at an earlier time.  By
recursively substituting Eq.~(\ref{eq:recursive}) into
Eq.~(\ref{eq:kab_phi}) the rate constant can be expressed as a product
of conditional probabilities:
\begin{eqnarray}
\label{rate4}
k_{AB} & = & \frac{ \left \langle \Phi_{A,\lambda_1} \right \rangle}{
\left \langle h_{\mathcal{A}}\right \rangle} \prod_{i=1}^{n-1}
\left\langle \bar{h}_{\Omega_{\lambda_{i+1}},A}^f \right
\rangle_{\Phi_{A,\lambda_i}} \left \langle \bar{h}_{B,A}^f \right
\rangle_{\Phi_{A,\lambda_n}} \nonumber\\ & \equiv & \frac{ \left
\langle \Phi_{A,\lambda_1} \right \rangle}{ \left \langle
h_{\mathcal{A}}\right \rangle} \prod_{i=1}^{n-1} {\mathcal
P}(\lambda_{i+1}| \lambda_{i}) {\mathcal P}(\lambda_{B} |\lambda_{n})
\\ & = & \frac{ \left \langle \Phi_{A,\lambda_1} \right \rangle}{
\left \langle h_{\mathcal{A}}\right \rangle} \left \langle
\bar{h}_{B,A}^f \right \rangle_{\Phi_{A,\lambda_1}} \equiv \frac{
\left \langle \Phi_{A,\lambda_1} \right \rangle}{ \left \langle
h_{\mathcal{A}}\right \rangle} {\mathcal P} (\lambda_{B} |\lambda_{1
}).\nonumber
\end{eqnarray}
This expression is the central equation for TIS.  Instead of just
calculating the individual terms in the product of Eq.~(\ref{rate4})
we can equivalently determine a continuous crossing probability
function ${\mathcal P}(\lambda |\lambda_{1})$ for $\lambda$ between
$\lambda_1$ and $\lambda_B$. This is reminiscent of umbrella sampling
where a free energy difference is usually estimated as a function of a
continuous parameter $\lambda$~\cite{FrenkelSmit}.  When calculating
the ensemble average for $ {\mathcal P}(\lambda_{i} |\lambda_{i-1}) $
we can also evaluate $ {\mathcal P}(\lambda |\lambda_{i-1}) $ for
interfaces $\lambda$ between $\lambda_{i-1}$ and $\lambda_i$ by
dividing the phase space into a finer grid of sub interfaces (see
fig.~(\ref{interfaces})). In this way we acquire useful information
without significant extra cost, and, in addition, a measure for the
convergence of the ensemble averages.  The final monotonically
decreasing crossing probability function ${\mathcal P}(\lambda
|\lambda_1)$ can be obtained by matching the histograms from the
different ensemble simulations. Techniques commonly applied in
umbrella sampling such as overlapping windows between two successive
ensemble averages and the use of biasing functions can also be
employed here.

\section{THE TRANSITION INTERFACE SAMPLING ALGORITHM}
\label{algorithm}
Inspection of Eq.~(\ref{rate4}) clearly shows that the TIS rate
constant calculation is also a two step procedure.  The first step,
the effective flux $\left \langle \Phi_{A,\lambda_1} \right \rangle/
\left \langle h_{\mathcal{A}}\right \rangle$ can be computed by simply
running a MD simulation starting with a configuration in $A$ and
counting the number of effective crossings. For an interface
$\lambda_1$ close enough to stable state $A$ one can obtain a
statistically accurate value.

The second part of the calculation consists of evaluating the product
of the ${\mathcal P}(\lambda_{i+1} |\lambda_{i})$ ensemble averages
for the different interfaces $\lambda_i$ in Eq.~(\ref{rate4}).  Here
we need to sample all paths from region $A$ to either $A$ or
$\Omega_{\lambda_{i+1}}$ that exhibit at least one crossing with
interface $\lambda_i$. The Monte Carlo moves in TIS are very similar
to the shooting move used in the TPS algorithm.  The main
difference is that the backward and forward integration is abandoned
as soon as the edge of either $A$ or $\Omega_{\lambda_{i+1}}$ is
reached.  If the new path is accepted there is only one phase point
$x$ along this path for which $\Phi_{A,\lambda_i}(x) \neq 0$, defining
phase space point $x_0$. The shifting moves that were required in the
original TPS implementation to enable proper sampling and improve
statistical accuracy are here unnecessary.

To bootstrap the sampling procedure we first generate an initial path
that starts in $A$, then crosses the interface $\lambda_i$ and finally
ends in either $A$ or $\Omega_{\lambda_{i+1}}$ (see for more details
on  initial path generation Ref.~\cite{Dellago02}).  The phase
space point $x_0$ is then defined as the first crossing point of this
path with interface $\lambda_i$. Further, let
$\tau=\textrm{int}(t/\Delta t)$ be the discrete time slice index, and
$\tau^b\equiv \textrm{int}(t_{A}^b(x_0)/\Delta t)$ and $\tau^f\equiv
\textrm{int}(t_{A\cup\Omega_{\lambda_{i+1}}}^f(x_0)/\Delta t)$ the
forward and backward terminal time slice indices, respectively.
Including $x_0$, the initial path then consists of
$N^{(o)}=\tau^b+\tau^f+1$ time slices.  With these definitions in mind
is the TIS algorithm as follows:

\begin{enumerate}
\item From the current path with length $N^{(o)}$ choose a random time
slice $\tau$, with $ -\tau^b \leq \tau \leq \tau^f$.
\item Change all momenta of $x_{\tau \Delta t}$ by adding small random
displacements $\delta p$ from a Gaussian distribution.  Make sure the
total momentum is conserved \cite{Dellago02}
\item\label{alg1} In case of a constant energy (NVE) simulation,
rescale the new momenta to the old energy value and continue with step
4.  In case of constant temperature (NVT) accept the new momenta (else
reject the whole TIS move) with a probability~\cite{FrenkelSmit}:
\begin{eqnarray}
\textrm{min} \Bigg[ 1,\exp\Big( \beta \big( E( x_{\tau \Delta
t}^{(o)}) -E(x_{\tau \Delta t}^{(n)}) \big) \Big) \Bigg]. \nonumber
\end{eqnarray}
Here, $E(x)$ is the total energy of the system at phase space point
$x$.

\item Integrate equations of motion backward in time by reversing the
momenta at time slice $\tau$, until reaching either $A$ or
$\Omega_{\lambda_{i+1}}$. Reject in case of $\Omega_{\lambda_{i+1}}$
else continue with the next step.

\item Integrate from time slice $\tau$ forward until reaching either
$A$ or $\Omega_{\lambda_{i+1}}$.  Reject if the entire trial path does
not cross interface $\lambda_i$, else continue with the next step.

\item\label{alg2} Accept the trial path with a probability
\begin{eqnarray}
\textrm{min} \Bigg[ 1,\frac{N^{(o)}}{N^{(n)}} \Bigg], \nonumber
\end{eqnarray}
where $N^{(n)}$ is the length of the new path. If accepted, replace
the old path with the new one.

\item Reassign $x_0$ to be the first crossing point with $\lambda_i$
and sample the value of $\bar{h}_{\Omega_{\lambda_{i+1}},A}^f(x_0)$ to
measure ${\mathcal P}(\lambda_{i+1}|\lambda_{i})$.

\item Repeat from step 1.
\end{enumerate}

As usual in Monte Carlo schemes, any rejection along this route
implies counting the old path again in the ensemble average.  The
acceptance probabilities at step \ref{alg1} and step \ref{alg2} are
required to satisfy the detailed balance condition (see
e.g. Ref~\cite{FrenkelSmit}).

Instead of generating a complete path and then accepting or rejecting
accordingly to the probability at step 6, it is more efficient to
determine a maximum path length in advance.  Before embarking on the
time consuming fourth and fifth step, we first take a uniform random
number $\alpha$ between $0$ and $1$ and determine the maximum allowed
path length by:
\begin{equation}
N_{\textrm{max}}^{(n)}=\textrm{int}(N^{(o)}/\alpha).
\end{equation}
In this way we can directly stop the integration and reject the TIS
move as soon the path length $N^{(n)}$ exceeds the maximum
$N_{\textrm{max}}^{(n)}$.  In the course of the TIS simulation the
path-length fluctuates. This also means that the average path length
becomes automatically shorter when changing from ensemble average
${\mathcal P}(\lambda_{i+1}|\lambda_{i})$ to ensemble average
${\mathcal P}(\lambda_{i}|\lambda_{i-1})$ closer to $A$.

The algorithm presented here does not require shifting moves because
there is only one unique $x_0$ phase point along each pathway.
However, one could consider the use of path-reversal moves as they
have negligible computational cost and can sometimes facilitate
ergodic sampling~\cite{Dellago02}.

\section{NUMERICAL RESULTS}
\label{application}

\subsection{The model}

We tested the TIS algorithm on a simple diatomic bistable molecule
immersed in a fluid of purely repulsive particles. Such a system has
previously been used in illustrating TPS rate constant calculations
\cite{TPS99} and is therefore a good starting point for a comparison
between the two methods.  The system consists of $N$ particles in 2
dimensions with interactions given by a pairwise Lennard-Jones (LJ)
potential truncated and shifted at the minimum, often referred to as
the Weeks-Chandler-Andersen (WCA) potential~\cite{WCA}
\begin{equation}
V_{WCA}(r)=
\begin{cases}
4\epsilon [(r/\sigma)^{-12}-(r/\sigma)^{-6}] + \epsilon & \mbox{if }
\, r\leq r_0 \\ 0 \quad \quad & \mbox{if } r>r_0, \\
\end{cases}
\end{equation}
where $r$ is the interatomic distance, and $r_0\equiv
2^{1/6}\sigma$. Throughout this section reduced units are used so that
$\epsilon$ and $\sigma$, respectively the LJ energy and length
parameters, as well as the mass of the particles are equal to
unity. The LJ unit of time $(m\sigma^2/\epsilon)^{1/2}$ is therefore
also unity.  In addition, two of the $N$ particles are interacting
through a double well potential
\begin{equation}\label{Vdw}
V_{dw}(r)=h\left[1-\frac{(r-r_0-w)^2}{w^2}\right]^2.
\end{equation}
This function has two minima separated by a barrier of height $h$
corresponding to the two stable states of the molecule: a compact
state for $r=r_0$ and extended state for $r=r_0+2w$. For a high enough
barrier, transitions between the states become rare and the rate
constant is well defined. Hence, this system provides a useful test
case for the TPS and TIS algorithms.

The system is simulated at a constant energy $E$ in a simulation
square box with periodic boundary conditions. The total linear
momentum is conserved and is set zero for all trajectories. The
equations of motion are integrated using the velocity Verlet algorithm
with a time step $\Delta t=0.002$. As in Ref.~\cite{TPS99} we focus
here on the computation of the rate constant for the isomerization
reaction of the dimer from the compact state to the extended state.
In the following section we describe general simulation details. In
Section~\ref{sec:HE} we discuss the results for a system with a high
enough barrier to avoid recrossings. Subsequently, we reproduce the
simulations from Ref.~\cite{TPS99} in Section~\ref{num.LE}. These
results do show recrossings, and we discuss the consequences for TPS
and TIS.

\subsection{Methodology}\label{num.met}

The TPS rate constant calculation evaluates the two factors in
Eq.~(\ref{kTPS}) separately as explained in Sec.~\ref{sec.TPS}. The
first term in Eq.~(\ref{kTPS}) is the ratio between the plateau value
of the reactive flux correlation function
$\langle\dot{h}_B(T)\rangle_{A,H_B(T)}$ and the correction $\langle
h_B(t') \rangle_{A,H_B(T)}$. The second term $C(t')$ requires an
umbrella sampling simulation in the form of a series of window
calculations.  An order parameter is chosen to define the
characteristic functions of the stable states and to partition phase
space in windows for the umbrella sampling.  Besides shooting and
shifting Monte Carlo moves to generate new paths in the transition
path sampling we also employ a \emph{diffusion} move that shifts the
path by one time slice in arbitrary direction.  This move is
computationally very cheap but increases the statistics of the
correlation functions.  In all our simulations we therefore set the
percentages for shooting, shifting and diffusion to 5\%,10\% and 85\%,
respectively.  The parameters involved are always gaged such that the
acceptance ratio is around 40\% for shooting and shifting moves,
ensuring an optimum efficiency of the sampling~\cite{TPS99}.

The TIS method involves a direct determination of the flux and the
calculation of the crossing probability functions ${\mathcal
P}(\lambda_{i}|\lambda_{i-1})$ between a series of successive
interfaces as given by Eq.~(\ref{rate4}).  The flux term in
Eq.~(\ref{rate4}) is computed by means of a straightforward MD
simulation starting in state $A$ and counting the number of effective
positive crossings through interface $\lambda_1$, i.e. when the
trajectory is directly coming from $A$.  The second term in
Eq.~(\ref{rate4}) is computed using the TIS algorithm of
Sec.~\ref{algorithm}. The basic requirement is a definition of a set
of interfaces partitioning the phase space. Between these interfaces
we defined a finer grid of sub-interfaces to construct the crossing
probability function ${\mathcal P}(\lambda|\lambda_{1})$.  As in the
TPS calculation we adjusted the momentum displacement for the shooting
move to give an acceptance of about 40\%.

Many parameters are involved in the two methods and to compare the
relative efficiency we measured the CPU-time required for an arbitrary
fixed error of $2.5\%$ for each step in both the TPS and TIS
calculations under the same computational conditions (1Ghz AMD
Athlon).  In both methods the final rate constant consists of a
product of factors which have to be calculated independently.  For
each factor we performed $M$ simulation blocks of $N$ Monte Carlo
cycles and adjusted $N$ such that after $M$ block averages the
relative standard deviation of each term in Eq.~(\ref{rateTPS}) and
(\ref{rateTIS}) was 2.5\%.  The total CPU-time is given by summing the
individual 2.5\% error CPU-times for each factor.  The final error in
the rate constants is obtained by the standard propagation rules using
all simulation results (i.e. not only the ones for the 2.5\% error CPU
time calculation).

\subsection{System with High Energy Barrier}\label{sec:HE}
\begin{figure}[t]
\begin{center}
\includegraphics[width=8cm,keepaspectratio]{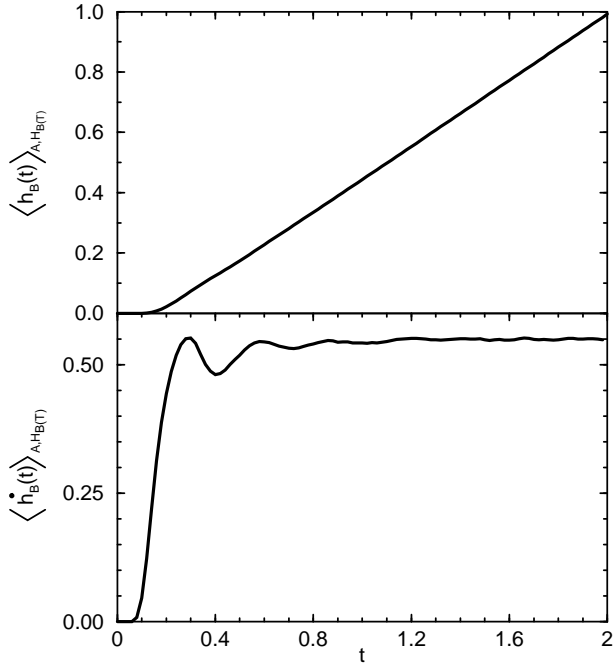}
\end{center}
\caption{TPS correlation function $\langle h_B(t)\rangle_{A,H_B(T)}$
(top) and its time derivative (bottom) for the system with high energy
barrier.  The error is comparable to line thickness.}
\label{TPS.corr.HE}
\end{figure}

\begin{figure}[t]
\begin{center}
\includegraphics[width=8cm,keepaspectratio]{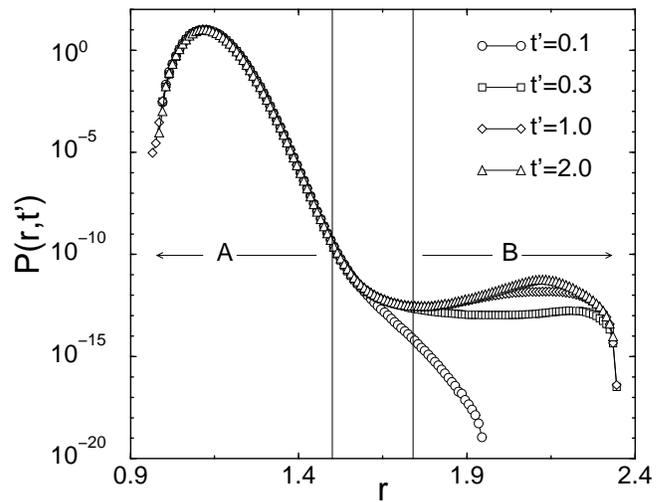}
\end{center}
\caption{TPS probability distributions $P(r,t')$ for four
$t'=0.1,0.3,1.0,2.0$ for the high energy barrier. The probability
$P(r,t')$ is the chance that a path of length $t'$ and starting in $A$
will have the end point conformation with a diatomic distance $r$. The
graph is the result of the matching of eight window
calculations. These eight windows are defined as $r<1.19$,
$1.18<r<1.28$, $1.27<r<1.35$, $1.34<r<1.40$, $1.39<r<1.47$,
$1.46<r<1.54$, $1.53<r<1.75$, $r>1.74$.  The errors on the histogram
points are within the symbol size.}
\label{TPS.umbr.HE}
\end{figure}

This system had a total number of particles $N=25$, and a total energy
$E=25$.  The square simulation box was adjusted to give a number
density of 0.7. The barrier height was $h=15$ and the width-parameter
$w=0.5$, so that the minima of $V_{dw}(r)$ were located at $r\simeq
1.12$ and $r\simeq 2.12$ while the top of the barrier was at $r\simeq
1.62$.  In the TPS rate calculation we defined stable states $A$ and
$B$ as $r<r_A=1.5$ and $r>r_B=1.74$, respectively. We computed the
correlation function $\langle h_B(t)\rangle_{A,H_B(T)}$ using TPS with
a fixed path length $T=2.0$.  The correlation function is shown in
Fig.  \ref{TPS.corr.HE} together with its time derivative, the
reactive flux.  The latter function clearly displays a plateau. Next,
we chose four different $t'=0.1, 0.3, 1.0, 2.0$ and performed umbrella
sampling simulations using 8 windows to calculate $C(t^\prime)$.  In
each window we measured the probability to find the path's end point
$r(t^\prime)$ at a certain value of $r$. These probability histograms
were rematched and normalized. The final probability functions are
shown in Fig.~\ref{TPS.umbr.HE}.  Integration of the area under the
histogram belonging to region B leads to $C(t')$ and finally to the
rate constant~\cite{TPS99}.  In Table \ref{restableHE} we give the
values of the different contributions to the rate constant given by
Eq.  \ref{kTPS}, together with the rate constant. We report the
average relative computation time needed to reach the $2.5\%$ error
(see Sec.~\ref{num.met}) in Table \ref{HE.MCmoves}.
\begin{table}[b]
\begin{center}
\begin{tabular}{c*{3}{r@{$\pm$}l}}
\multicolumn{7}{l}{TPS} \\ \hline \hline $t'$ &
\multicolumn{2}{c}{$\frac{\langle\dot{h}_B(T)\rangle_{AB}} {\langle
h_B(t')\rangle_{AB}}$} & \multicolumn{2}{c}{$C(t')/10^{-13}$} &
\multicolumn{2}{c}{$k_{A\rightarrow B}/10^{-13}$} \\ 0.1 & 3300&100 &
0.0018&0.0001 & 6.0&0.5 \\ 0.3 & 7.54&0.03 & 0.76&0.02 & 5.8&0.1 \\
1.0 & 1.236&0.005 & 4.8&0.3 & 5.9&0.4\\ 2.0 & 0.553&0.002 & 11.4&0.9 &
6.3&0.5 \\ \multicolumn{7}{c}{} \\ \multicolumn{7}{l}{TIS} \\ \hline
\hline & \multicolumn{2}{c}{$ \langle \Phi_{A,\lambda_1} \rangle
/{\langle h_{\mathcal{A}}\rangle}$} & \multicolumn{2}{c} {${\mathcal
P}(\lambda_B|\lambda_1)/10^{-13}$} &
\multicolumn{2}{c}{$k_{A\rightarrow B}/10^{-13}$} \\ & 0.1196&0.0005 &
49&1 & 5.9&0.2
\end{tabular}
\end{center}
\caption{Comparison of rate constants for the high energy barrier,
computed with TPS at different $t'$ and TIS.  Contributing factors
from Eq.~(\ref{kTPS}) and Eq.~(\ref{rate4}) are also given.}
\label{restableHE}
\end{table}

\begin{figure}[t]
\begin{center}
\includegraphics[width=8cm,keepaspectratio]{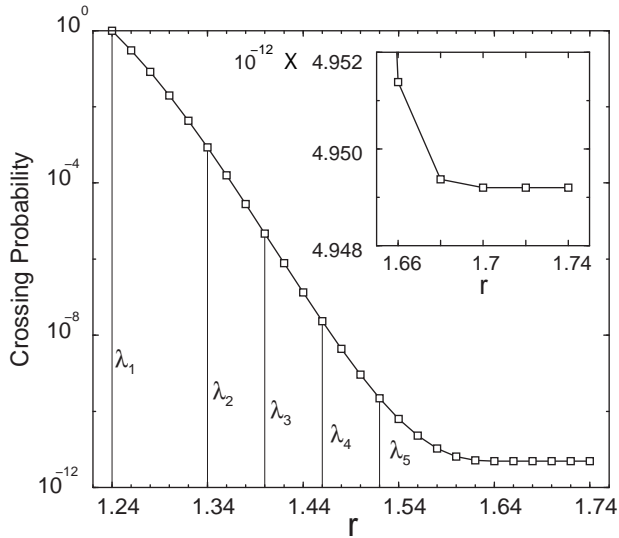}
\end{center}
\caption{TIS crossing probability ${\mathcal P}(\lambda|\lambda_1)
=\langle \bar{h}_{\Omega_ {\lambda},A}^f \rangle_{\Phi_{A,\lambda_1}}$
as function of $\lambda=r$ for the system with a high energy barrier.
The function is computed by matching the five interface ensemble
calculations. These interfaces were chosen at:
$\lambda_1=1.24,\lambda_2=1.34,\lambda_3=1.40, \lambda_4=1.46$ and
$\lambda_5=1.52$.  The error on the points is within symbol size.  The
inset is an enlargement in linear scale of the last part of the
function.  We clearly detect a horizontal plateau when approaching
$\lambda_B$.}
\label{HE.TISB}
\end{figure}
\begin{table}[b]
\begin{center}
\begin{tabular}{*{11}{c}}
\multicolumn{11}{l}{TPS} \\ \hline \hline $t'$ &
$\frac{\langle\dot{h}_B(T)\rangle_{AB}}{\langle h_B(t')\rangle_{AB}}$&
W1 & W2 & W3 & W4 & W5 & W6 & W7 & W8 & Total\\ $0.1$ & 11.0 & 0.01 &
0.05 & 0.1 & 0.04 & 0.23 & 0.27 & 1.3 & 0.01 & 13.01 \\ $0.3$ & 0.2 &
0.01 & 0.14 & 0.28 & 0.13 & 0.58 & 0.43 & 0.19 & 0.02 & 1.98 \\ $1.0$
& 0.1 & 1.7 & 1.7 & 0.9 & 0.6 & 3.0 & 2.6 & 6.4 & 0.2 & 17.2 \\ $2.0$
& 0.1 & 0.03 & 1.8 & 4.5 & 4.4 & 15.3 & 8.0 & 20.3 & 0.6 & 55.03 \\
\end{tabular}\\
\begin{tabular}{*{7}{c}}
\multicolumn{7}{l}{TIS} \\ \hline \hline ${ \langle \Phi_{A,\lambda_1}
\rangle}/{ \langle h_{\mathcal{A}} \rangle}$ & Int $\lambda_1$ & Int
$\lambda_2$ & Int $\lambda_3$ & Int $\lambda_4$ & Int $\lambda_5$ &
Total time \\ 0.07 & 0.265 & 0.09 & 0.15 & 0.21 & 0.215 & 1\\
\end{tabular}
\end{center}
\caption{Comparison of CPU-times required for the 2.5\% error at each
stage for the system with high energy barrier. The times are
renormalized to the TIS total computation time. W1 to W8 denote the
different windows used in the calculation, Int $\lambda_1$ to Int
$\lambda_5$ denote the interface ensemble calculations.}
\label{HE.MCmoves}
\end{table}


%
For the TIS calculations we use the same order parameter $r$ and the
same definition for region $B$, i.e. interface $\lambda_{B}$ is set at
$r=1.74$.  Stable state $A$ was defined by setting
$\lambda_A=\lambda_1$ at $r=1.24$.  This interface is closer to the
basin of attraction than the TPS stable state definition but yields a
higher flux term $\langle \Phi_{A,\lambda_1} \rangle / \langle
h_{\mathcal{A}} \rangle$ and gives better statistics.  Note that the
different definition of stable state $A$ does not change the final
rate constant, as the overall state $\mathcal{A}$ does not sensitively
depends on this definition.  The flux term is calculated by
straightforward NVE MD.  As $\lambda_A$ is equal to $\lambda_1$ every
positive crossing of this interface is counted in the flux because all
trajectories must by default come directly from A.  The conditional
crossing probabilities ${\mathcal P}(\lambda_{i+1}|\lambda_{i})$ in
Eq.~(\ref{rate4}) are calculated for $n=5$ interfaces between the
stable states (see fig. \ref{HE.TISB}).  Between these interfaces we
impose a finer grid to obtain the entire crossing probability
function.  The results for each stage and the final rate constant are
shown in Table \ref{restableHE}.  The rate constants of both methods
agree within the statistical accuracy, showing that the TIS method is
correct.  In Table \ref{HE.MCmoves} we give the relative computation
time to reach the $2.5\%$ error for each term.

\begin{figure}[t]
\begin{center}
\includegraphics[width=8cm,keepaspectratio]{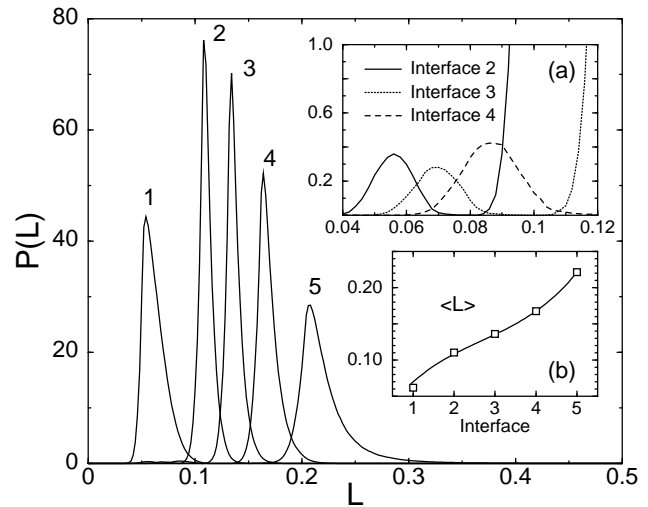}
\end{center}
\caption{Histograms $P(L)$ of path length $L$ for each ensemble,
computed for the system with the high energy barrier.  Inset (a) is an
enlargement of the bottom left area, where windows 2,3,4 display a
second peak. They represent that small fraction of paths that are able
to cross all the interfaces up to the rightmost interface and do not
have to return to A (cf. the trajectories with the white circle in
Fig.3).  Inset (b): average path length in each window. At variance
with TPS the TIS algorithm adapts the path length to the ensemble.  In
going from interface 5 to interface 1 one gets closer to state $A$ and
the path length shortens accordingly.}
\label{TIS.plength}
\end{figure}
In comparing both methods we have to recall that the efficiency of TPS
depends strongly on the choice of $t'$.  On the one hand the umbrella
calculation of $C(t')$ is much faster for low values of $t'$. But on
the other hand the error in the correction term $\langle h_B(t')
\rangle_{A,H_B(T)}$ increases for lower $t'$.  As a result there is an
optimum $t'$ for the error/CPU-time ratio, in this case approximately
at $t'=0.3$.  Even for this optimized situation the TIS calculation is
about two times faster.  One could object that the correlation
function in Fig. \ref{TPS.corr.HE} has reached a plateau for $t=1.5$
already, reducing the TPS computation time by a factor $3/4$.  But the
choice for a path length $T=1.5$ can not be taken without a-priori
knowledge.  The first term in Eq.~(\ref{kTPS}) implicitly depends on
the path length $T$.  Changing $T$ would alter the ensemble and might
result in a different shape of the flux correlation function.  We did
not check this in detail, but we believe that $T$ cannot be chosen
much smaller without introducing systematic errors.  Furthermore, we
emphasize here that we put much more effort in optimizing the TPS
algorithm by tuning $t'$, the windows, the ratio between shooting,
shifting and diffusion moves than we did for TIS.

Figure~\ref{TIS.plength} shows the histograms of path lengths for each
TIS ensemble calculation and shows why TIS is faster.  Sampling paths
of fixed length with TPS results in spending unnecessary computation
time inside the initial and final stable regions A and B.  In the TIS
algorithm instead every path is adapted to its minimum
length. Bringing the interface in closer to $A$ reduces these
transition times. TIS optimizes itself during the simulation.

\begin{figure}[t]
\begin{center}
\includegraphics[width=8cm,angle=0,keepaspectratio]{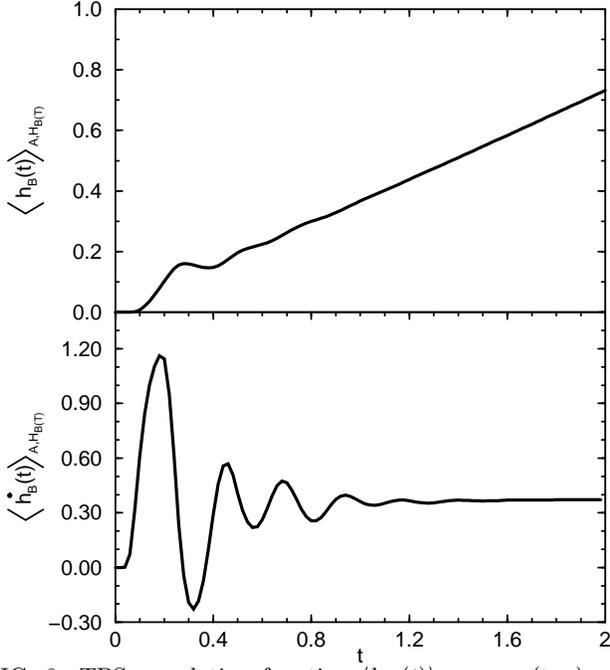}
\end{center}
\vspace{-1cm}
\caption{TPS correlation function $\langle h_B(t)\rangle_{A,H_B(T)}$
(top) and its time derivative (bottom) for the system with low energy
barrier.  The error is comparable to line thickness.}
\label{LETPScorr}
\end{figure}

\subsection{System with Low Energy Barrier}\label{num.LE}
In order to compare with previous results, we adopted the parameters
from Ref.~\cite{TPS99}.  The total number of particles was $N=9$, the
total energy was $E=9$ and the square simulation box was adjusted for
a number density of 0.6. The barrier height is $h=6$ and the
width-parameter is $w=0.25$. Minima are at $r\simeq 1.12$ and $r\simeq
1.62$, while the top of the barrier is at $r\simeq 1.37$.  This
barrier is much lower than in the previous section resulting in more
frequent transitions. An approximate rate constant could even be
achieved by straightforward MD simulations.

\begin{table}[b]
\begin{center}
\begin{tabular}{c*{3}{r@{$\pm$}l}}
\multicolumn{7}{l}{TPS} \\ \hline \hline $t'$ &
\multicolumn{2}{c}{$\frac{\langle\dot{h}_B(T)\rangle_{AB}} {\langle
h_B(t')\rangle_{AB}}$} & \multicolumn{2}{c}{$C(t')/10^{-5}$} &
\multicolumn{2}{c}{$k_{A\rightarrow B}/10^{-5}$} \\ 0.1 & 47.3&0.2 &
1.408&0.007 & 6.67&0.04 \\ 0.4 & 2.505&0.007 & 2.67&0.01 & 6.68&0.03\\
0.8 & 1.240&0.003 & 5.42&0.05 & 6.72&0.07 \\ 2.0 & 0.507&0.001 &
13.9&0.2 & 7.03&0.09 \\ \multicolumn{7}{c}{} \\
\multicolumn{7}{l}{TIS} \\ \hline \hline & \multicolumn{2}{c}{$\frac{
\langle \Phi_{A,\lambda_1} \rangle}{\langle h_{\mathcal{A}}\rangle}$}
& \multicolumn{2}{c}{$\langle
\bar{h}_{B,A}^f\rangle_{\Phi_{A,\lambda_1}}/10^{-5}$} &
\multicolumn{2}{c}{$k_{A\rightarrow B}/10^{-5}$} \\ & 0.2334&0.0003 &
29.6&0.2 & 6.90&0.06
\end{tabular}
\end{center}
\caption{ Comparison of rate constants for the low energy barrier
computed with TPS at different $t'$ and with TIS, including the
contributing factors from Eq.~(\ref{kTPS}) and Eq.~(\ref{rate4}),
respectively. Computation times are reported in units of the TIS
CPU-time.}
\label{restableLE}
\end{table}
\begin{figure}[t]
\begin{center}
\includegraphics[height=8cm,angle=-90,keepaspectratio]{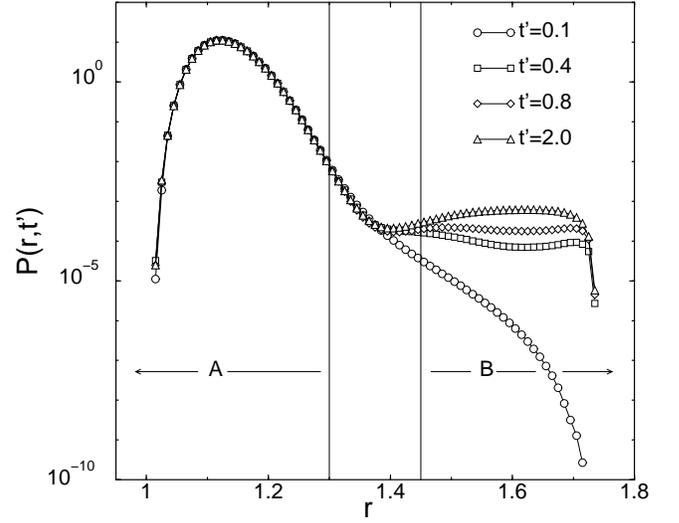}
\end{center}
\vspace{-0.5cm}
\caption{TPS probability distributions $P(r,t')$ for four
$t'=0.1,0.3,1.0,2.0$ for the system with low energy barrier $P(r,t')$
as in Fig.~\ref{TPS.umbr.HE}. The graph is the result of the matching
of five window calculations.  These five window calculations are
defined as $r<1.22$, $1.21<r<1.26$, $1.25<r<1.30$, $1.29<r<1.46$,
$r>1.45$ The errors on the histogram points are within the symbol
size.}
\label{LETPSumbr}
\end{figure}

\begin{table}[b]
\begin{center}
\begin{tabular}{*{8}{c}}
\multicolumn{8}{l}{TPS} \\ \hline \hline $t'$ &
$\frac{\langle\dot{h}_B(T)\rangle_{AB}}{\langle h_B(t')\rangle_{AB}}$&
W1 & W2 & W3 & W4 & W5 & Total \\ $0.1$ & 0.68 & 0.03 & 0.009 & 0.01 &
0.1 & 0.001 & 0.83\\ $0.4$ & 0.4 & 0.09 & 0.03 & 0.04 & 0.25 & 0.01 &
0.82\\ $0.8$ & 0.28 & 0.21 & 0.07 & 0.11 & 1.5 & 0.04 & 2.21\\ $2.0$ &
0.35 & 0.28 & 0.38 & 0.93 & 7.27 & 0.14 & 9.35\\
\end{tabular}\\
\begin{tabular}{*{5}{c}}
\multicolumn{5}{l}{TIS} \\ \hline \hline $\frac{ \langle
\Phi_{A,\lambda_1} \rangle}{ \langle h_{\mathcal{A}} \rangle}$ & Int
$\lambda_1$ & Int $\lambda_2$ & Int $\lambda_3$ & Total \\ 0.015 &
0.085 & 0.45 & 0.45 & 1\\
\end{tabular}
\end{center}
\caption{Comparison of CPU-times required for the 2.5\% error at each
stage for the system with the low energy barrier. The times are
renormalized to the TIS total computation time.}
\label{LE.MCmoves}
\end{table}

\begin{figure}[t]
\begin{center}
\includegraphics[width=8cm,angle=0,keepaspectratio]{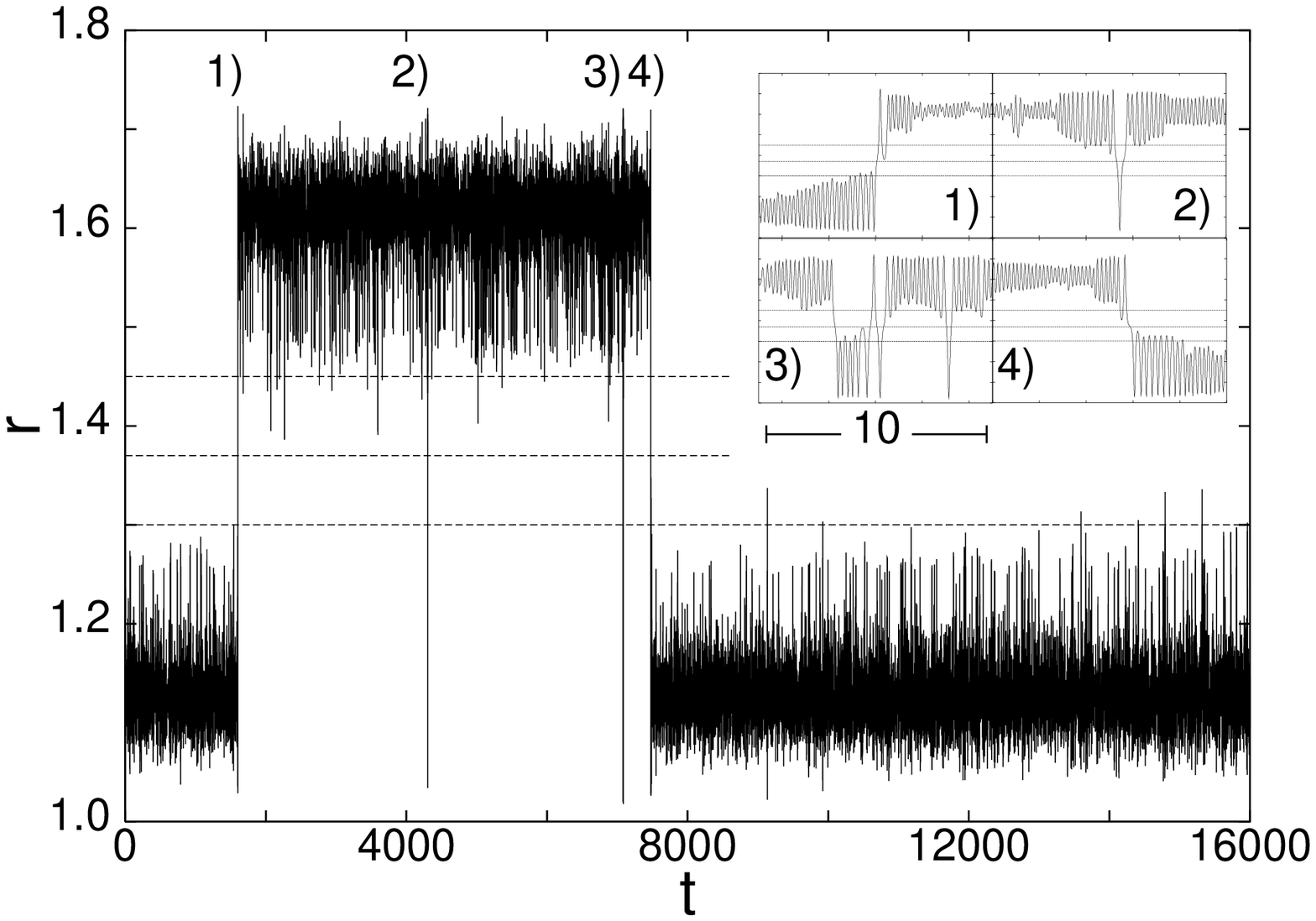}
\end{center}
\caption{Intra-molecular distance of the dimer as function of time
from a straightforward MD simulation for the system with the low
energy barrier.  Horizontal dashed line at 1.37 corresponds to the top
of the potential barrier. Horizontal dashed lines at 1.3 and 1.45
correspond to the TPS state definitions of
Ref.~\protect\cite{TPS99}. Insets are enlargements of four typical
events on a scale of $10$.  1) and 4) correspond to true reactive
events, $A\rightarrow B$ and $B\rightarrow A$ respectively while 2)
and 3) are non-true, fast recrossing events.  In particular, event 3)
shows capricious behavior with many crossings of the barrier.  The
figure shows a clear separation of timescales, $t_{\textrm{mol}} \sim
1$ and $t_{\textrm{rxn}} \sim 1000$.}
\label{yoyos}
\end{figure}

For the TPS calculations we defined the stable states $A$ and $B$ by
$r<r_A=1.30$ and $r>r_B=1.45$, respectively~\cite{TPS99}.  Using
standard TPS simulation we computed the correlation function $\langle
h_B(t)\rangle_{A,H_B(T)}$ with a total path length $T=2$ (shown in
Fig.~\ref{LETPScorr}).  Next, we measured the probability histograms
to find the paths end point at a certain order parameter value $r$ for
four different times $t'=0.1, 0.4, 0.8, 2.0$, using five
windows~\cite{TPS99} (see Fig.~\ref{LETPSumbr}).  As described in the
previous section, matching the probability histograms and subsequent
integration leads to $C(t')$. The resulting final rate constants,
shown in Table \ref{restableLE}, are comparable with the results of
Ref. \cite{TPS99}, but more accurate.  We will discuss these values
after giving the results of TIS.

\begin{figure}[t]
\begin{center}
\includegraphics[height=8cm,angle=-90,keepaspectratio]{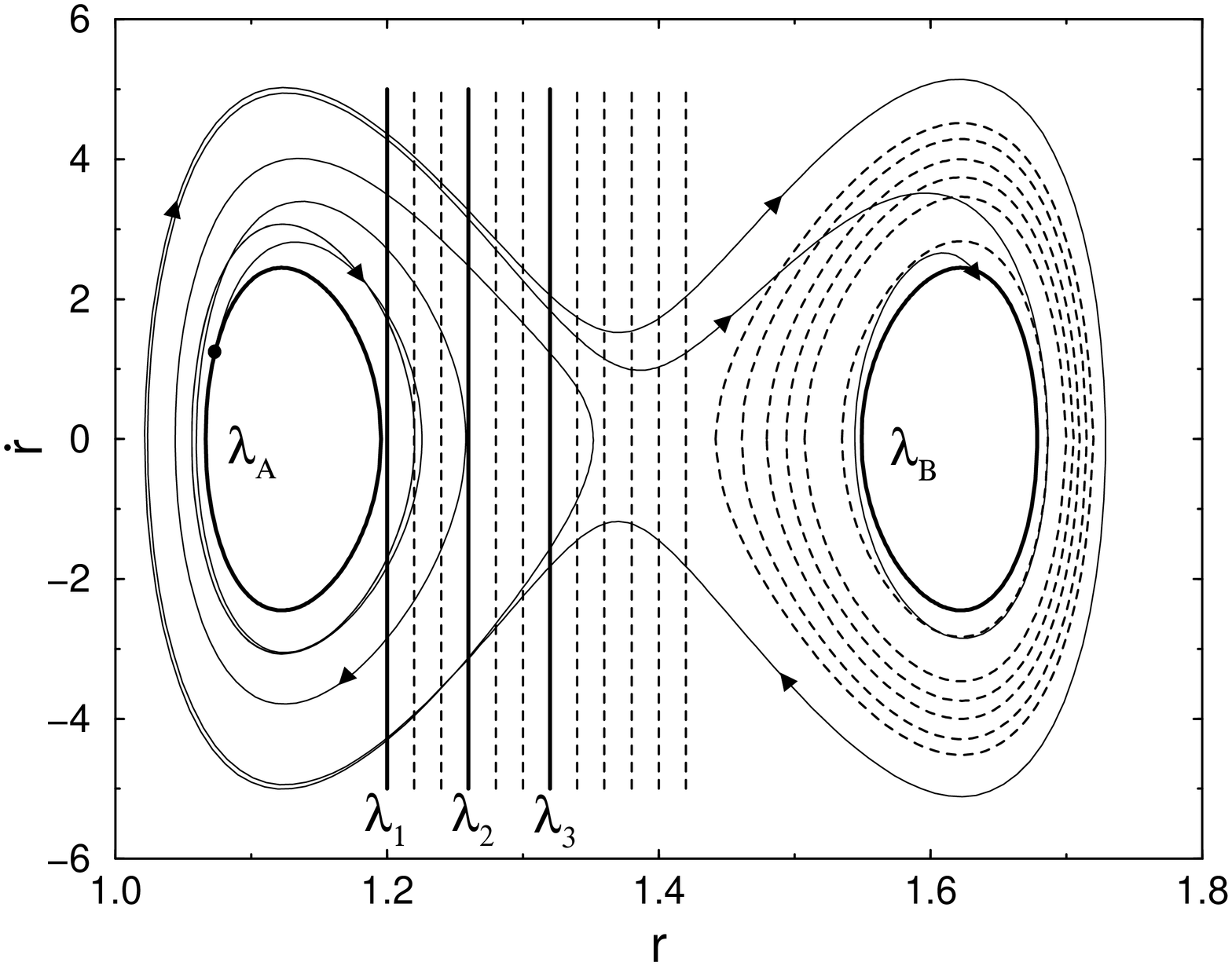}
\end{center}
\caption{One calculated path of the low energy barrier system shown in
the $\{r,\dot{r} \}$ plane. The vertical solid lines are the interface
$\lambda_1$,$\lambda_2$ and $\lambda_3$.  The curves $\lambda_A$ and
$\lambda_B$ are the boundaries of the TIS stable states. The dashed
lines are the sub-interfaces.  The path starts at the dot on
$\lambda_A$ and crosses the barrier three times before dissipating its
energy and relaxing into state $B$.  } \label{interfacesLE}
\vspace{-3mm}
\end{figure}
\begin{figure}[!hb]
\begin{center}
\includegraphics[width=8cm,keepaspectratio]{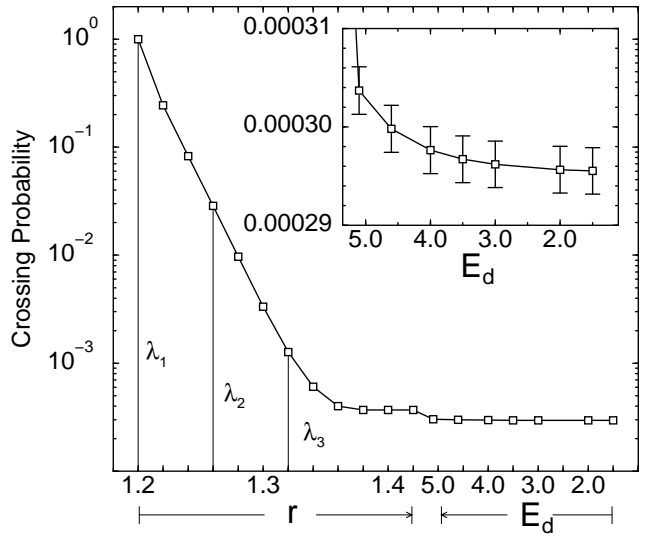}
\end{center}
\caption{The crossing probability ${\mathcal P}(\lambda|\lambda_1)$
for the system with the low energy barrier.  The function is computed
by matching ensemble calculations with interfaces $\lambda_1$ at
$r=1.20$, $\lambda_2$ at $r=1.26$ and $\lambda_3$ at $r=1.32$.  The
inset is an enlargement of the final part.  The function is converging
to a plateau but has not yet reached it. The different values of the
last points are due to the presence of fast recrossings.The error is
inside the symbol size.}
\label{LE.TISB}
\end{figure}
Figure~\ref{yoyos} shows that fast recrossings can occur for a low
barrier, implying that $r$ alone is not sufficient as an order
parameter to define the stable states in the simulations. Apparently,
this does not effect the TPS results much, but it is very important
for TIS because of the assumption that stable region B is really
stable and recrossings do not take place.  To ensure the stability of
the TIS stable states we chose a new order parameter that not only
depends on the inter-atomic distance $r$ in the dimer but also on a
kinetic term, given by $\dot{r}$.  The stable states can then be
defined by
\begin{eqnarray}\label{interf.eq}
E_d(r,\dot{r})\equiv\frac{\dot{r}}{4}+V_{dw}(r)\nonumber \\ x \in A
\textrm{ if } r < 1.37 \textrm{ and } E_d(r,\dot{r}) \leq 1.5
\nonumber \\ x \in B \textrm{ if } r > 1.37 \textrm{ and }
E_d(r,\dot{r}) \leq 1.5,
\label{defABLE}
\end{eqnarray}
where $E_d$ is the sum of the kinetic and potential energy of the
dimer that has a reduced mass of $1/2$.  In the $\{r,\dot{r}\}$ plane
these stable states form a D-shape and an inverse D-shape regions for
$A$ and $B$ respectively (see Fig.~\ref{interfacesLE}).  Crossing the
interface $\lambda_A$ or $\lambda_B$ implies that the vibrational
energy is decreased below the threshold, $E_d=1.5$.  This threshold is
made low enough to make fast recrossings to the other state
unlikely. However, if we would have chosen it too low the paths would
have become very long.  We evaluated the crossing probability function
in Eq.~(\ref{rate4}) for $n=3$ interfaces.  The entire crossing
probability function was obtained by partitioning the phase space in
sub-interfaces of the form $r=\lambda$ and $E_d(r,\dot{r})=\lambda$ as
shown in Fig.~\ref{interfacesLE}.  Note that in TIS multidimensional
or multiple order parameters can be used in one simulation without a
problem. This is more difficult in TPS, where a proper mapping of the
complete phase space is required.  Figure \ref{LE.TISB} shows the
final rematched crossing probability. The monotonically decreasing
function tends to reach a plateau on approaching the last interface.
The last two values are not exactly equal but differ by 0.03\%,
indicating that a small fraction of the paths crossing the one but
last sub-interface still succeed to return to $A$ without crossing
$\lambda_B$.  This difference is comparable with the chance of a new
independent transition (given by the rate constant).  Note that
without the kinetic energy definition for the stable states
Eq.~(\ref{defABLE}), the final crossing probability and thus the rate
constant would have been overestimated by a factor $5/4$.

For the effective flux $\left \langle \Phi_{A,\lambda_1} \right
\rangle/ \left \langle h_{\mathcal{A}}\right \rangle$ calculation we
performed MD simulations as described in Sec~\ref{num.met}.  In
contrast to the high barrier case, $\lambda_1$ is not equal
$\lambda_A$, and not all positive crossings with $\lambda_1$ are
effective crossings. We counted only the first crossing when the
system left region $A$ and waited until the system fell back to region
$A$ before counting a new crossing.  As the MD trajectory sometimes
displayed a spontaneous transition to region $B$, we stopped the
simulation and started again by replacing the system in a randomized
configuration of $A$.  Table~\ref{restableLE} shows the final values
and the corresponding errors of these calculations.  The relative
computation time for each term is detailed in table~\ref{LE.MCmoves}.

\begin{figure}
\begin{center}
\includegraphics[height=8cm,angle=-90,keepaspectratio]{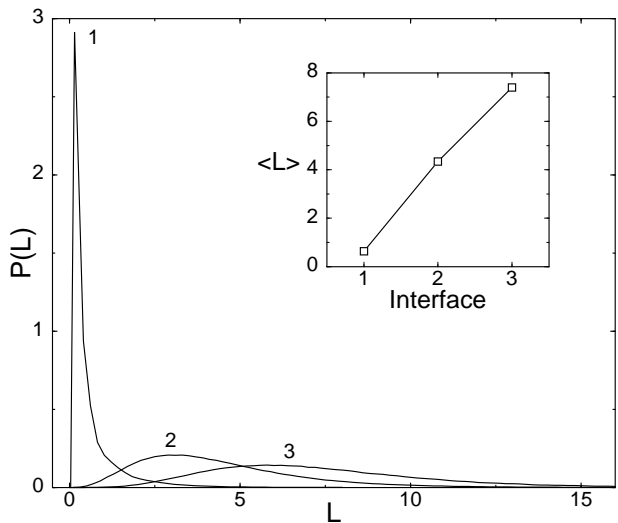}
\end{center}
\caption{Path length distribution $P(L)$ for each interface ensemble
in the low energy barrier system The inset shows the average path
length in each ensemble.}
\label{LEhisto}
\end{figure}

If we compare the final results of table~\ref{restableLE} we see that
the efficiency of TIS is more than nine times better than the TPS
efficiency for $t'=2$, and more than two times better than TPS value
for $t'=0.8$. But the TPS $t'=0.1$ and $t'=0.4$ efficiencies are about
20 \% better than TIS.  When we compare the rate constants, however,
we notice that the TPS results for different $t'$ do not agree.  Among
the TPS rate constants only the $t'=2$ case is consistent with the TIS
result.  We believe that the $t'=0.1$ and $t'=0.4$ results suffer from
systematic errors.  For instance, for the shorter paths the TPS
simulations might not be completely ergodic.  Another explanation
might be that a path length of $T=2$ is too short to allow convergence
of the reactive flux.  In the TIS calculation the average path length
in the three interface simulations, from the closest to $B$ to the
closest to $A$, is, respectively, 7.4, 4.3, and 0.63; much longer than
the TPS path length (see Fig.\ref{LEhisto}).  It is therefore
surprising that the TPS approach with the simple stable state
definition and very short paths still gives approximately the right
rate constant.  And indeed, when we computed the TPS correlation
function with the TIS state definitions Eq.~(\ref{defABLE}), we found
that the path length had to be at least $T=20$ to see a plateau.  We
think that TPS works even with the simple state definitions and the
short paths because both positive flux and negative flux terms
contribute to Eq.~(\ref{kTPS}).  The TPS algorithm collects many paths
of which some are not real transitions, but fast recrossings.  The
cancellation of positive and negative terms of these fast recrossing
paths ensure the (almost) correct final outcome. In TIS each path must
be true transition event and contributes as a positive term in rate
equation (\ref{rate4}), enhancing the convergence.  This explains that
the CPU time for the TIS calculation despite the much longer paths is
still comparable with TPS one for low $t'$.  We note that the path
ensemble using the more strict stable state definition is of course
more useful in the analysis of the reaction mechanism.

For a more accurate comparison of the computation time we must keep
the systematic errors lower than the statistical errors. In other
words, we have to make sure that the results are converged.  To test
the convergence of the flux correlation function in TPS we can derive
the following equality from Eq.~(\ref{kTPS}) :
\begin{equation}
\frac{ \langle h_B(t') \rangle_{A,H_B(T)} }{ \langle h_B(t'')
\rangle_{A,H_B(T)} }= \frac{ C(t') }{ C(t'') }.
\label{convTest}
\end{equation}
This equation is valid for any $t',t'' < T$ if $T$ is large enough.
We found that the equality does not hold for the system with the low
barrier, indicating that $T$ is too low in the TPS calculation.
Further examination of the flux correlation function $\langle h_B(t')
\rangle_{A,H_B(T)}$ reveals that the apparent plateau has in fact a
small positive slope. Calculations for higher values of $T$ suggest
that one has to increase the path length at least to $T=8$ to
convergence to a plateau.  With this in mind we think that the TIS
computation is about a factor five more efficient than the TPS
algorithm for the model system with the low barrier.

\section{CONCLUSION}
\label{conclusion}
We developed a novel method, named transition interface sampling, for
the calculation of rate constants based on transition path sampling
concepts. Just as the original transition path sampling, the new
method enables the calculation of rate constants of transitions
between stable states separated by high free energy barriers without
prior knowledge of the reaction coordinate. The new algorithm is
different in spirit from the rate constant calculation that was
introduced in Refs\cite{TPS98a, TPS99}. In TPS the time correlation
function $\langle h_A(x_0) h_B(x_t) \rangle/\langle h_A \rangle$ is
determined for a single time using an umbrella sampling scheme
followed by a calculation of the reactive flux prefactor in a separate
path sampling simulation. The path length used in this simulation has
to be long enough for the plateau to be reached.  The TIS method
advocated here calculates the flux correlation directly by measuring
the fluxes through a number of different interfaces and relating the
flux through one interface to the next one.  The big advantage of a
flux instead of a correlation function is that trajectories going
through the interfaces all contribute to the rate whereas in TPS there
are recrossings to be counted.  In addition, the new method improves
the original TPS method on a several other points.  Once the interface
is reached, the integration of motion can be stopped instead of going
all the way to region $B$.  In this way the TIS algorithm adapts
itself to the optimal path length.  One does not have to optimize the
new method as much as TPS, where one has to find the optimal $t'$
value and a proper balance between shooting and shifting.  Besides
being faster, the concept of calculating a flux comes natural with the
rate constant definition, and implementation of the algorithm is hence
simpler.  Also, multidimensional or even discrete order parameters can
easily be implemented in TIS.  In the illustrative example we showed
that we can obtain an increase in efficiency of at least a factor of
two to five with respect to the TPS method used in Ref.~\cite{TPS99}.

However, one has to be more careful in the definitions of the stable
states, meaning that stable states have to be really stable.  In TPS
the choice of stable states is a bit more flexible as the final rate
constant consists of cancellation of positive and negative terms.  In
sec \ref{num.LE} we showed how this problem for TIS can be solved by
defining stable regions that explicitly depend on kinetic energy
terms.

In summary, we believe that the TIS algorithm can make the rate
constant calculation of many processes feasible that were hitherto
difficult to obtain.  For instance, chemical reactions in solution,
isomerization of clusters and conformational transitions in
biomolecules. In a future publication we will report on these, more
complex, applications.

\acknowledgements T.S.v.E acknowledges NWO-CW (Nederlandse Organisatie
voor Wetenschappelijk Onderzoek, Chemische Wetenschappen).  P.G.B
acknowledges support from the FOM (Stichting Fundamenteel Onderzoek
der Materie). We thank Evert Jan Meijer for carefully reading 
the manuscript.

\begin{appendix}
\section{FLUX RELATION} 
In this appendix we show how the effective flux $\Phi_{A,\lambda_i}$
can be related to the effective flux $\Phi_{A,\lambda_{i-1}}$ through
an interface $\lambda_{i-1}$ closer to $A$. If at time $t=0$ a
trajectory passes interface $\lambda_i$ while having started in $A$
some time earlier, there must always be an unique time when it passed
interface $\lambda_{i-1}$ for the first time.  Therefore we can write:
\begin{eqnarray}
\Phi_{A,\lambda_i}(x_0)= \Phi_{A,\lambda_i}(x_0) \int_0^{t_{A\cup
\Omega_{\lambda_i}}^b(x_0)} \mathrm{d} t \,
\Phi_{A,\lambda_{i-1}}(x_{-t})
\end{eqnarray}
and hence,
\begin{eqnarray}
\left \langle \Phi_{A,\lambda_i}(x_0) \right \rangle = \nonumber \\
\int_0^{\infty} \mathrm{d} t \left \langle
\Phi_{A,\lambda_{i-1}}(x_{-t}) \Phi_{A,\lambda_i}(x_0) \theta( t_{
A\cup \Omega_{\lambda_i} }^b(x_0) -t) \right \rangle
\nonumber\nonumber \\ = \int_0^{\infty} \mathrm{d} t \left \langle
\Phi_{A,\lambda_{i-1}}(x_{0}) \Phi_{A,\lambda_i}(x_t) \theta( t_{A\cup
\Omega_{\lambda_i}}^b(x_t) -t) \right \rangle \nonumber \\ = \left
\langle \Phi_{A,\lambda_{i-1}} (x_{0}) \int_0^{\infty} \mathrm{d} t
\Phi_{A,\lambda_i}(x_t) \theta(t_{A\cup \Omega_{\lambda_i}}^b(x_t)-t)
\right \rangle \nonumber\nonumber \\ =\left \langle
\Phi_{A,\lambda_{i-1}}(x_0) \int_0^{t_{A \cup \Omega_{\lambda_i}}^f
(x_0)} \mathrm{d} t \Phi_{A,\lambda_i}(x_t) \right \rangle\nonumber \\
= \left \langle \Phi_{A,\lambda_{i-1}}(x_0)
\bar{h}_{\Omega_{\lambda_i},A}^f(x_0) \right \rangle.
\label{eqn15}
\end{eqnarray}
The one but last equation follows because for each phase point $x$ and
phase space region $\Omega$ it can be shown that $t> t_\Omega^f (x)
\Rightarrow t_\Omega^b( f(x,t)) \leq t \Rightarrow \theta\bigg(
t_\Omega^b( f(x,t)) -t\bigg)=0$.  We rewrite the last expression of
Eq.~(\ref{eqn15}) as a different ensemble average:
\begin{eqnarray}
\left \langle \Phi_{A,\lambda_{i-1}}(x_0)
\bar{h}_{\Omega_{\lambda_i},A}^f(x_0) \right \rangle=\nonumber \\
=\frac{\left \langle \Phi_{A,\lambda_{i-1}}(x_0)
\bar{h}_{\Omega_{\lambda_i},A}^f(x_0) \right \rangle} {\left \langle
\Phi_{A,\lambda_{i-1}}(x_0) \right \rangle } \times {\left \langle
\Phi_{A,\lambda_{i-1}}(x_0) \right \rangle } \nonumber\\ \equiv \left
\langle \bar{h}_{\Omega_{\lambda_i},A}^f(x_0) \right
\rangle_{\Phi_{A,\lambda_{i-1}}} \times \left \langle
\Phi_{A,\lambda_{i-1}}(x_0) \right \rangle ,
\end{eqnarray}
where $\left \langle \ldots \right \rangle_{\Phi_{A,\lambda_{i-1}}}$
denotes the ensemble average over all phase points $x_0$ for which
$\Phi_{A,\lambda_{i-1}}(x_0) \neq 0$. The last equality gives rise to
Eq.~(\ref{eq:recursive}).

\end{appendix}

\newpage


\newpage

\newpage

\end{document}